\documentclass[useAMS,usenatbib]{mn2e}
\usepackage{graphicx}
\usepackage{txfonts}
\usepackage{fixltx2e}

\title[Asteroseismology of KUV~08368+4026]{Asteroseismology of the ZZ Ceti star KUV~08368+4026\thanks{Based on data obtained
at the Xinglong station of National Astronomical Observatories of China,
the Lijiang station of Yunnan Astronomical Observatory, China,
the San Pedro Martir Observatory, Mexico,
the Observatorio Astrofisico Guillermo Halo, Mexico,
and the Observatoire de Haute Provence, France.}}
\author[Li et al.]
{C. Li$^1$; J.-N. Fu$^1$\thanks{Send offprint request to: J.-N. Fu(jnfu@bnu.edu.cn)}; G. Vauclair$^{2,3}$; N. Dolez$^{2,3}$; L. Fox-Machedo$^4$; R. Michel$^4$ and M. Chavez$^5$ \\
$^1$Department of Astronomy, Beijing Normal University, Beijing, China\\
$^2$Universit\'{e} de Toulouse; UPS-OMP; IRAF; Toulouse, France\\
$^3$CNRS; IRAF; 14 avenue Edouard Belin, F-31400, Toulouse; France\\
$^4$Observetorio Astro\'{o}mico Nacional, Instituto de Astronomia, Universided Nacional Aut\'{o}noma de M\'{e}xico, Ensenada, BC, M\'{e}xico\\
$^5$Instituto Nacional de Astrof\'{i}sics \'{O}ptica y Electr\'{o}nica (INAOE), Tonantzintla, Pue, Mexico}

\begin{document}

\date{Accepted; Received}

\pagerange{\pageref{firstpage}--\pageref{lastpage}} \pubyear{2002}

\maketitle

\label{firstpage}

\begin{abstract}
Asteroseismology is a unique tool to explore the internal structure of stars
through both observational and theoretical research. The internal structure of
pulsating hydrogen shell white dwarfs (ZZ Ceti stars) detected by
asteroseismology is regarded as the representative of all DA white dwarfs.
Observations for KUV~08368+4026, which locates in the middle of the ZZ Ceti
instability strip, have been carried out in 1999 and from 2009 to 2012 with either single-site runs or multisite campaigns. Time-series photometric data of about 300 hours were collected in total. Through data
reduction and analysis, 30 frequencies were extracted, including four
triplets, two doublets, one single mode and further signals. The independent modes are identified as
either l=1 or l=2 modes. Hence, a rotation period of $5.52\pm 0.22$ days was
deduced from the period spacing in the multiplets. Theoretical static models
were built and a best fit model for KUV~08368+4026 was obtained with $0.692\pm0.002$ solar mass, $(2.92\pm0.02)\times 10^{-3}$ solar
luminosity and the hydrogen mass fraction of $10^{-4}$ stellar mass.
\end{abstract}


\section{Introduction}
White dwarfs are the final remains of almost all the moderate- and low-mass stars and the oldest kind of stars in the Galaxy. Measurement of the age of white dwarfs can put constraints on the galaxy's and the universe's ages(Winget et al. 1987). The accurate determination of age requires precise measurements of stellar parameters such as the total mass, luminosity, radius, effective temperature, hydrogen mass fraction, helium mass fraction and so on. Asteroseismology provides a tool to estimate these parameters through modeling the internal structure of pulsating white dwarfs.

Since the discovery of the first member by Landolt (1968), pulsating white dwarfs are divided into four classes (DAV, DBV, DQV and GW Vir). Among them, the DAVs or ZZ Ceti stars or hydrogen surface pulsating white dwarfs have the lowest effective temperatures and the largest total number of members. The instability strip of ZZ Ceti stars locates at the cross of the Cepheid instability strip and the evolution tracks of white dwarfs. It is regarded as a ``pure'' instability strip, which indicates that every DA white dwarf locating in the instability strip does show pulsation. Thus the internal structure of ZZ Ceti stars can be regarded as the representative of all DA white dwarfs.

For the ZZ Ceti stars, the properties vary dependent on their locations in the instability strip. The hot ZZ Ceti stars locating close to the blue edge of the instability strip exhibit typically short periods, low amplitudes and small amplitude modulations, while the cold members show usually long periods, high amplitudes and large amplitude modulations. The attempt of using theoretical models to explore the internal structure DA white dwarfs has been applied to either individual ZZ Ceti stats(c. f. HL Tau 76, Pech, Vauclair \& Dolez 2006; HS 0507 Fu et al. 2013) or globally for a sample of ZZ Ceti stars(Romero et al. 2012).

KUV~08368+4026 was discovered to be a ZZ Ceti star by Vauclair et al. (1997). A three site campaign was made in 1998 by Dolez et al. (1999). Fontaine et al.(2003) gave stellar parameters including the effective temperature of 11490 K, the surface gravity log g of 8.05, the mass of 0.64 solar mass and the absolute magnitude of 11.85. However, a set of parameters was given by Gianninas et al. (2011), providing a hotter model with $T_{eff}$ of $12280\pm192K$, log g of $8.17\pm0.05$, mass of $0.71\pm0.03M_{\sun}$ and absolute magnitude of 11.88. In order to study the oscillations of KUV~08368+4026 hence constrain the stellar parameters, one-week observations were made from Haute-Provence Observatory in 1999 and four observation runs were taken from 2009 to 2012 from multiple site.+

The observation and data reduction are described in section 2. We present the period and seismology analysis in section 3 and 4, respectively. Section 5 gives discussion on the variations of pulsation amplitudes in different time scales. In section 6, we present the attempt of stellar modeling and the result. Finally we give the discussion and conclusions in section 7.

\section{Observations and data reduction}
Time-series photometric data (dataset 1) were collected for KUV~08368+4026 withe the Chevreton photoelectric photometer in 1999 from Haute-Provence Observatory in France. Four runs were taken for this star from 2009 to 2012 when CCD cameras and Johnson B filters were used. The run of dataset 2 was a single site observation effort carried out from Lijiang Observatory in China in February of 2009 with the 2.4m telescope. From December of 2009 to January of 2010, data (dataset 3) were obtained with the 2.4-m telescope in Lijiang and the 2.16-m telescope in Xinglong of China, and the 1.5-m telescope in Observatorio de San Pedro M\'{a}rtir (SPM) of Mexico. In the run of dataset 4, two more telescopes in Xinglong (80-cm and 85-cm telescopes) were used, together with the 2.16-m telescope in Xinglong and the 2.1-m telescope in Observatorio Astrofisico Guillermo Halo (OAGH) in Mexico. The run of dataset 5 was arranged as a two site campaign of the 2.16m-telescope in Xinglong and the 1.5-m telescope in SPM. Unfortunately, observations in Mexico were not carried out due to some technical problems.

Table 1 lists the observation log. All data were reduced with the package of IRAF DAOPHOT except the photometer's data, which were reduced with the standard method for photoelectric photometer data. Figure 1 shows the reduced light curves of KUV~08368+4026.

\begin{table}
 \centering
  \caption{Journal of observations for KUV~08368+4026}
  \begin{tabular}{@{}ccccc@{}}
    \hline\noalign{\smallskip}
Dataset & Date & Observatory & Telescope & Frame \\
  \hline\noalign{\smallskip}
1 & Jan.19-25,1999 & Haute-Provence &1.93m & - \\
 \hline   2 & Feb.10-18,2009 & Lijiang & 2.4m & 6805  \\
\hline ~& Dec.12-18,2009 & Xinglong & 2.16m & 1651 \\
~& Dec.26-31,2009 & Lijiang & 2.4m & 2328 \\
3& Dec.27-28,2009 & Xinglong & 2.16m &815  \\
~& Jan.07-17,2010 & SPM & 1.5m &2771 \\
~& Jan.12-19,2010 & Xinglong & 2.16m & 2745\\
\hline ~& Mar.01-03,2011 &Xinglong & 80cm &1516\\
~& Mar.04-07,2011& Xinglong & 85cm &1746\\
4& Mar.08-10,2011& Xinglong & 2.16m &1655\\
~& Mar.08-11,2011& OAGH & 2.1m &1556\\
\hline 5 & Feb.17-23,2012 & Xinglong & 2.16m & 3976 \\
\hline
\end{tabular}
\end{table}

\begin{figure*}
\resizebox{\hsize}{!}{\includegraphics[angle=90]{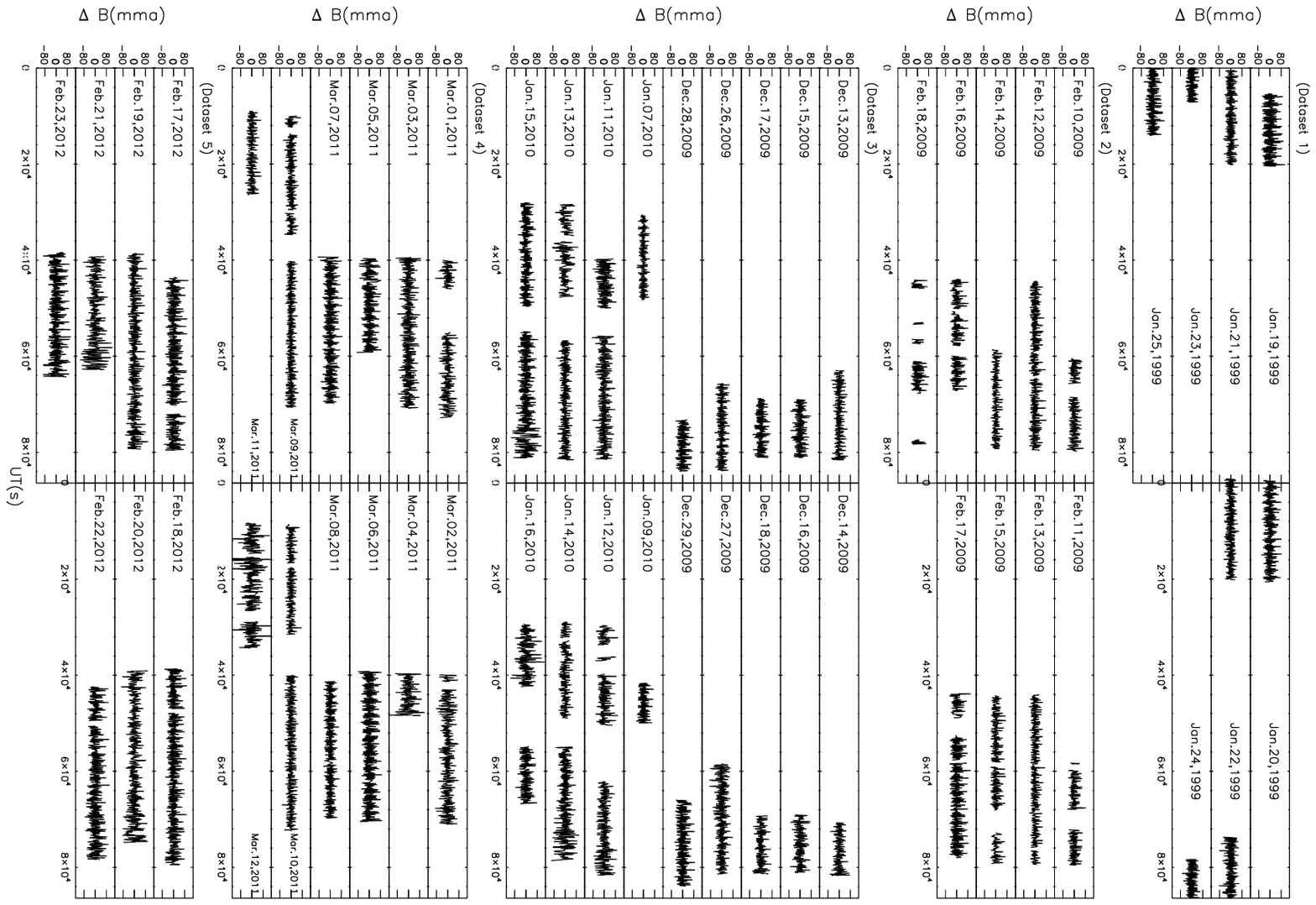}}
\caption{Light curves of all our observations in B for KUV~08368+4026.}
\label{fig1}
\end{figure*}

\section{Period analysis}

Period analysis was made for the light curves by using the software {\tt Period04} (Lenz \& Breger 2005). For all the five datasets, we made Fourier transformation. Figure 2 shows the Fourier spectra. Notice that the amplitudes of the same frequencies vary among the observation runs and this will be discussed in Section 5.

\begin{figure}
\resizebox{\hsize}{!}{\includegraphics[angle=90]{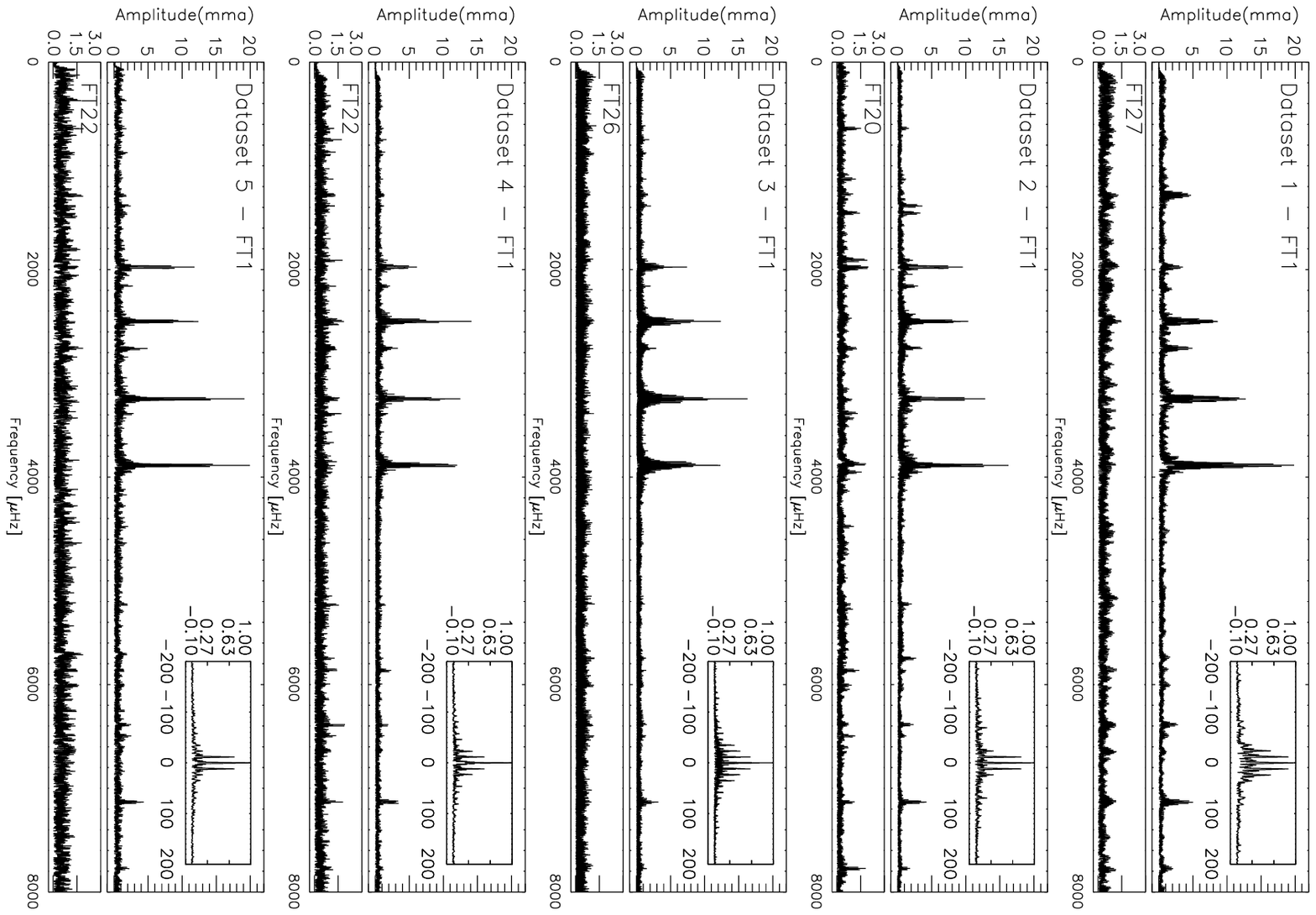}}
\caption{FTs of the light curves for the five datasets from the top to the bottom. For each dataset the full spectrum before prewhitening is shown in the upper panel while the spectrum window and the residuals are shown in the insets and the lower panels, respectively.}
\label{fig2}
\end{figure}

The data analysis followed such steps: 1) extract the highest peak from the Fourier spectrum, get its frequency and amplitude, and calculate its phase through fitting. 2) Prewhiten the sine function, make the Fourier transformation for the residuals in order to get the next frequency. We repeated the steps and got finally a list of frequencies with S/N ratio larger than 4. The extracted frequencies are listed in Table 2. The method of Monte-Carlo simulation was used to derive the uncertainties. For more detail about Monte-Carlo simulation please read Fu et al. 2013.

\begin{table*}
\setlength\tabcolsep{2.5pt}
\centering
\small
\caption{Frequency list for the five datasets. Note that $f$ is the frequency in $\mu Hz$ and A is the amplitude in mma. The ``Note'' column indicates the linear combinations, $ia=11.574~\mu Hz$ is the one-day aliasing. The label of the frequencies for each dataset follows the order of decreasing amplitude.}                                    
\begin{tabular}{@{}lrrc|lrrc|lrrc|}
\hline  \multicolumn{4}{c}{(Dataset 1)}&\multicolumn{4}{c}{(Dataset 2)}& \multicolumn{4}{c}{(Dataset 3)}\\
 ID & $f$ & A & Note & ID & $f$ & A & Note & ID & $f$ & A & Note   \\
\hline
  a24 &  747.31$\pm$0.22 &  1.56$\pm$0.73 & a2-a4&   b9 & 1386.90$\pm$0.05 &  3.63$\pm$0.61 & b1-b6&  c16 &  747.51$\pm$0.08 &  1.84$\pm$0.66 & c1-c2 \\
  a26 & 1115.80 $\pm$0.33&  1.55$\pm$0.67 & &  b11 & 1454.67$\pm$0.08 &  3.36$\pm$0.59 & &  c22 &  875.50$\pm$0.09 &  1.42$\pm$0.72 & \\
  a16 & 1278.72$\pm$0.21 &  2.44$\pm$0.80 & &   b4 & 1976.48$\pm$0.03 &  9.92$\pm$0.65 & &  c20 & 1127.55$\pm$0.13 &  1.53$\pm$0.59 & c4-c11\\
  a10 & 1280.51$\pm$0.14 &  4.23$\pm$0.78 & &  b13 & 2159.09$\pm$0.07 &  2.75$\pm$0.63 & &  c15 & 1270.98$\pm$0.04 &  1.85$\pm$0.75 & c1-c21\\
  a9 & 1283.64$\pm$0.13 &  5.24$\pm$0.77 & &   b3 & 2498.76$\pm$0.02 &  9.71$\pm$0.62 & &  c13 & 1387.15$\pm$0.06 &  2.27$\pm$0.71 & c4-c2\\
  a21 & 1961.68$\pm$0.53 &  1.70$\pm$0.82 & a14-a9&   b6 & 2500.71$\pm$0.02 &  8.03$\pm$0.70 & &  c21 & 1975.43$\pm$0.05 &  3.07$\pm$0.91 & \\
  a12 & 1976.10$\pm$0.15 &  3.32$\pm$1.05 & &  b19 & 2757.87$\pm$0.08 &  2.12$\pm$0.58 & &   c5 & 1976.44$\pm$0.02 &  7.64$\pm$0.90 &  \\
  a20 & 2191.49$\pm$0.39 &  1.80$\pm$0.94 & &  b10 & 2759.99$\pm$0.05 &  3.34$\pm$0.64 & &  c12 & 1977.42$\pm$0.05 &  3.16$\pm$0.65 & \\
  a6 & 2498.85$\pm$0.05 &  7.33$\pm$0.87 & &  b16 & 3245.51$\pm$0.11 &  2.73$\pm$0.52 & &  c14 & 2159.25$\pm$0.08 &  2.21$\pm$0.72 & \\
  a4 & 2501.15$\pm$0.07 &  7.42$\pm$0.71 & &   b2 & 3246.74$\pm$0.03 & 11.95$\pm$0.72 & &   c2 & 2498.91$\pm$0.01 & 12.60$\pm$0.77 &  \\
  a19 & 2606.27$\pm$0.58 &  1.77$\pm$0.99 & &   b8 & 3247.77$\pm$0.06 &  4.85$\pm$0.69 & &   c6 & 2500.62$\pm$0.03 &  5.90$\pm$0.96 & \\
  a8 & 2758.81$\pm$0.10 &  5.12$\pm$0.79 & &  b17 & 3249.11$\pm$0.11 &  2.50$\pm$0.58 & b1+b2-b5&  c11 & 2758.44$\pm$0.04 &  2.74$\pm$0.67 & \\
  a22 & 2760.75$\pm$0.28 &  1.74$\pm$0.80 & a18-a3&   b5 & 3885.53$\pm$0.02 & 10.09$\pm$0.47 & &  c10 & 2760.33$\pm$0.08 &  2.97$\pm$0.82 & \\
  a14 & 3244.98$\pm$0.25 &  3.24$\pm$0.90 & &  b12 & 3886.75$\pm$0.04 &  4.35$\pm$0.85 & &   c1 & 3246.39$\pm$0.01 & 15.41$\pm$0.98 & \\
  a3 & 3246.43$\pm$0.10 &  9.83$\pm$0.83 & &   b1 & 3887.73$\pm$0.02 & 15.31$\pm$0.76 & &  c25 & 3247.29$\pm$0.03 &  2.40$\pm$0.97 & c9-c3 \\
  a2 & 3248.13$\pm$0.08 & 12.91$\pm$0.73 & &  b18 & 5223.79$\pm$0.07 &  2.17$\pm$0.61 & b4+b8&   c8 & 3248.56$\pm$0.02 &  5.39$\pm$0.82 & \\
  a11 & 3249.32$\pm$0.24 &  4.24$\pm$0.93 & &  b14 & 5747.68$\pm$0.07 &  2.74$\pm$0.70 & b6+b2&   c4 & 3886.01$\pm$0.01 & 11.12$\pm$0.78 &  \\
  a1 & 3886.25$\pm$0.02 & 20.68$\pm$0.77 & &  b15 & 6386.52$\pm$0.06 &  2.40$\pm$0.60 & b3+b1&   c3 & 3887.13$\pm$0.02 &  9.76$\pm$0.91 & \\
  a5 & 3888.25$\pm$0.06 &  7.35$\pm$0.62 & &   b7 & 7134.37$\pm$0.05 &  4.28$\pm$0.70 & b2+b1&   c7 & 3888.02$\pm$0.02 &  8.28$\pm$0.74 & \\
  a23 & 4526.86$\pm$0.73 &  1.59$\pm$0.77 & a16+a2&  & & & &  c23 & 5747.48$\pm$0.09 &  1.43$\pm$0.72 & c6+c1\\
  a25 & 5735.79$\pm$0.33 &  1.54$\pm$0.82 & &  & & & &  c19 & 5863.48$\pm$0.06 &  1.48$\pm$0.61 & c12+c4\\
  a18 & 6007.03$\pm$0.24 &  1.87$\pm$0.88 & a8+a2&  & & & &  c18 & 6385.99$\pm$0.13 &  1.64$\pm$0.74 & c2+c3\\
  a13 & 6387.46$\pm$0.10 &  3.04$\pm$0.64 & a4+a1&  & & & &  c24 & 6492.81$\pm$0.09 &  1.41$\pm$0.75 & \\
  a17 & 6495.03$\pm$0.33 &  1.93$\pm$0.72 & a3+a2&  & & & &   c9 & 7134.27$\pm$0.03 &  3.32$\pm$0.69 & c1+c7 \\
  a7 & 7134.46$\pm$0.08 &  5.27$\pm$0.85 & a2+a1&  & & & &  c17 & 7774.06$\pm$0.08 &  1.66$\pm$0.87 & c4+c7\\
  a15 & 7774.32$\pm$0.17 &  2.29$\pm$0.81 & a1+a5&  & & & &  & & & \\
\hline
\multicolumn{4}{c}{(Dataset 4)}& \multicolumn{4}{c}{(Dataset 5)}& \multicolumn{4}{c}{ }\\
 ID & $f$ & A & Note & ID & $f$ & A & Note &   &   &   &    \\
\hline
  d17 & 1975.23$\pm$0.49 &  2.68$\pm$1.60 & &  e21 &  872.54$\pm$0.25 &  2.08$\pm$1.20 &   & & & &\\
   d6 & 1976.33$\pm$0.22 &  7.87$\pm$2.89 & &  e13 & 1387.16$\pm$0.19 &  2.71$\pm$1.47 & e1-e3   & & & &\\
   d9 & 1977.21$\pm$0.35 &  4.92$\pm$3.78 & &  e22 & 1454.68$\pm$0.34 &  2.05$\pm$1.14 &   & & & &\\
  d21 & 2159.12$\pm$0.36 &  2.07$\pm$0.92 & &  e15 & 1911.02$\pm$0.68 &  2.32$\pm$1.31 & e1-e4  & & & &\\
  d12 & 2488.34$\pm$0.14 &  3.46$\pm$0.92 & d1-ia&   e4 & 1976.51$\pm$0.06 & 12.06$\pm$1.34 &   & & & &\\
   d1 & 2498.94$\pm$0.06 & 14.15$\pm$0.98 & &  e11 & 2159.33$\pm$0.28 &  2.69$\pm$1.52 &   & & & &\\
  d10 & 2509.96$\pm$0.12 &  4.11$\pm$0.86 & d1+ia&   e5 & 2498.65$\pm$0.09 &  9.47$\pm$1.17 &   & & & &\\
  d16 & 2514.12$\pm$0.16 &  2.67$\pm$0.92 & &   e3 & 2500.66$\pm$0.06 & 10.69$\pm$1.45 &   & & & &\\
   d18 & 2748.33$\pm$0.35 &  2.30$\pm$0.77 & d14-ia&   e8 & 2760.33$\pm$0.14 &  4.79$\pm$1.37 &   & & & &\\
  d14 & 2758.30$\pm$0.27 &  3.22$\pm$0.83 & &   e2 & 3246.46$\pm$0.04 & 18.69$\pm$1.45 &   & & & &\\
   d7 & 3245.72$\pm$0.14 &  6.21$\pm$2.26 & &   e7 & 3248.44$\pm$0.10 &  6.90$\pm$1.32 &   & & & &\\
   d2 & 3246.79$\pm$0.22 & 14.55$\pm$1.84 & &  e10 & 3387.63$\pm$0.18 &  3.18$\pm$1.42 &   & & & &\\
   d8 & 3247.96$\pm$0.53 &  5.82$\pm$1.80 & &   e6 & 3885.75$\pm$0.09 &  9.16$\pm$1.52 &   & & & &\\
  d15 & 3874.58$\pm$0.55 &  4.94$\pm$0.96 & d5-ia&   e1 & 3887.78$\pm$0.04 & 18.02$\pm$1.40 &   & & & &\\
  d11 & 3885.73$\pm$0.19 &  7.15$\pm$5.69 & &  e18 & 3952.66$\pm$0.26 &  2.22$\pm$1.28 &   & & & &\\
   d3 & 3886.81$\pm$0.21 & 17.06$\pm$2.28 & &  e20 & 4477.19$\pm$0.20 &  2.10$\pm$1.13 & e4+e3   & & & &\\
  d5 & 3888.04$\pm$0.23 &  8.64$\pm$1.60 & &  e19 & 5234.48$\pm$0.29 &  2.21$\pm$1.08 &   & & & &\\
   d4 & 3897.58$\pm$0.27 &  6.30$\pm$1.67 & d3+ia&  e12 & 6386.16$\pm$0.26 &  2.61$\pm$1.20 & e5+e1   & & & &\\
  d20 & 5863.43$\pm$0.41 &  2.13$\pm$0.78 & d6+d3&  e14 & 6494.87$\pm$0.20 &  2.55$\pm$1.27 & e2+e7  & & & &\\
  d19 & 7122.88$\pm$0.35 &  2.64$\pm$1.09 & d8+d15&  e16 & 7132.20$\pm$0.23 &  2.34$\pm$1.17 & e2+e6  & & & &\\
  d13 & 7145.87$\pm$0.53 &  2.52$\pm$0.85 & d8+d4&   e9 & 7134.29$\pm$0.16 &  4.26$\pm$1.34 & e2+e1  & & & &\\
  & & & &  e17 & 7773.71$\pm$0.51 &  2.23$\pm$1.1019 & e6+e1\\
\hline
\end{tabular}
\end{table*}

For dataset 3, we found a number of extremely-close frequencies with low amplitudes in addition to the major frequencies during the prewhitenning process. After checking the data carefully we realized that it is due to changing amplitudes of the same frequencies in the time scale of weeks, thus the prewhitenning of the frequency could not be completely done with a fixed amplitude and phase. The subroutine ``Calculate amplitude/phase variations'' of Period04 was then used to solve this problem. For each frequency, individual amplitude and phase were used to prewhiten the light curves of idividual run. The residuals were then combined to calculate the next Fourier transformation.

We compare frequency list of the five datasets in Table 2 to each other in order to identify the same frequencies from different datasets. The result is presented in Table 3.

\begin{table*}
\centering
\caption{Comparison of the fequencies and amplitudes in the five datasets. Note that $f$=frequency in $\mu$Hz, $A$=amplitude in mma. }
\begin{tabular}{@{}lrrrrrrrrrr}
\hline  & \multicolumn{2}{c}{(Dataset 1)}&\multicolumn{2}{c}{(Dataset 2)}& \multicolumn{2}{c}{(Dataset 3)}& \multicolumn{2}{c}{(Dataset 4)}& \multicolumn{2}{c}{(Dataset 5)}\\
ID & $f$ & A & $f$ & A &  $f$ & A & $f$ & A& $f$ & A\\

 \hline
 F1&         &       &        &       &        &       &        &       &   872.54 &  2.08 \\
 F2&         &       &        &       &   875.50 &  1.42 &        &       &        &       \\
 F3&   1115.80 &  1.55 &        &       &        &       &        &       &        &       \\
 F4&   1278.72 &  2.44 &        &       &        &       &        &       &        &       \\
 F5&   1280.51 &  4.23 &        &       &        &       &        &       &        &       \\
 F6&   1283.64 &  5.24 &        &       &        &       &        &       &        &       \\
 F7&         &       &  1454.67 &  3.36 &        &       &        &       &  1454.68 &  2.05 \\
 F8&         &       &        &       &  1975.43 &  3.07 &  1975.23 &  2.68 &        &       \\
 F9&   1976.10 &  3.32 &  1976.48 &  9.92 &  1976.44 &  7.64 &  1976.33 &  7.87 &  1976.51 & 12.06 \\
F10&         &       &        &       &  1977.42 &  3.16 &  1977.21 &  4.92 &        &       \\
F11&         &       &  2159.09 &  2.75 &  2159.25 &  2.21 &  2159.12 &  2.07 &  2159.33 &  2.69 \\
F12&   2191.49 &  1.80 &        &       &        &       &        &       &        &       \\
F13&   2498.85 &  7.33 &  2498.76 &  9.71 &  2498.91 & 12.60 &  2498.94 & 14.15 &  2498.65 &  9.47 \\
F14&   2501.15 &  7.42 &  2500.71 &  8.03 &  2500.62 &  5.90 &        &       &  2500.66 & 10.69 \\
F15&         &       &        &       &        &       &  2514.12 &  2.67 &        &       \\
F16&   2606.27 &  1.77 &        &       &        &       &        &       &        &       \\
F17&   2758.81 &  5.12 &  2757.87 &  2.12 &  2758.44 &  2.74 &  2758.30 &  3.22 &        &       \\
F18&         &       &  2759.99 &  3.34 &  2760.33 &  2.97 &        &       &  2760.33 &  4.79 \\
F19&  3244.98 &  3.24&  3245.51 &  2.73 &        &       &  3245.72 &  6.21 &        &       \\
F20&   3246.43 &  9.83 &  3246.74 & 11.95 &  3246.39 & 15.41 &  3246.79 & 14.55 &  3246.46 & 18.69 \\
F21&   3248.13 & 12.91 &  3247.77 &  4.85 &  3248.56 &  5.39 &  3247.96 &  5.82 &  3248.44 &  6.90 \\
F22&   3249.32 &  4.24 &        &       &        &       &        &       &        &       \\
F23&         &       &        &       &        &       &        &       &  3387.63 &  3.18 \\
F24&   3886.25 & 20.68 &  3885.53 & 10.09 &  3886.01 & 11.12 &  3885.73 &  7.15 &  3885.75 &  9.16 \\
F25&         &       &  3886.75 &  4.35 &  3887.13 &  9.76 &  3886.81 & 17.06 &        &       \\
F26&   3888.25 &  7.35 &  3887.73 & 15.31 &  3888.02 &  8.28 &  3888.04 &  8.64 &  3887.78 & 18.02 \\
F27&         &       &        &       &        &       &        &       &  3952.66 &  2.22 \\
F28&         &       &        &       &        &       &        &       &  5234.48 &  2.21 \\
F29&   5735.79 &  1.54 &        &       &        &       &        &       &        &       \\
F30&         &       &        &       &  6492.81 &  1.41 &        &       &        &       \\

\hline
\end{tabular}
\end{table*}
\section{Asteroseismology}
\subsection{Linear combinations}
Since most of runs are single site observations, aliasing effect is strongly visible in some spectrum windows in Figure 2, especially in the window spectra of January of 1999 (dataset 1), Febuary of 2009 (dataset 2) and Febuary of 2012 (dataset 5).

The analysis of linear combinations and aliasing frequencies were made and the result is listed in the column "Note" of Table 2.

\subsection{Mode identification}
After removing the frequencies of linear combination and aliasing, we list the frequencies with amplitudes in Table 3. The frequencies which are components of multiplets, or present large amplitudes, or are detected in multiple seasons are identified as independent signals. For the single frequencies with low amplitudes or detected in only one observing season, we group them as further signals. Please note the frequency of 3249.32~$\mu Hz$. Although it is close to a triplet, we assign it as a further signal since it is detected only once in a single-site observing run with a photoelectric photometer and its amplitude is small. We summarize the independent frequencies and further signals in Table 4. For the frequencies detected in multiple seasons, we take the average values of the frequencies.

Four triplets are identified from the independent signals in Table 4. f1-f3 is an unequal triplet. Since the spacing of f3 to f2 is almost twice of the spacing of f1 to f2, we suppose they are l=2 modes. The other three triplets show nearly equal spacing of about 1~$\mu Hz$ inside the triplets, which is the property of l=1 modes. We also notice that the two doublets have frequency spacing of about 2~$\mu Hz$. Thus we identify them as l=1 modes with the central frequencies of m=0 mode undetected.

From the equation:
\[
\sigma_{k,l,m}=\sigma_{k,l}+m\times (1-C_{k,l}) \Omega
\]

Where $C_{k,l}=1/l(l+1)$ in the asymptotic regime (Brickhill 1975), one may derive that the splits of the l=2 modes are 1.67 times of the splits of the l=1 modes. As far as the triplet around 1280.51~$\mu Hz$, the two splits are about 1.6 and 3.2
~$\mu Hz$, which agrees with the early identification of this frequency as a l=2 mode.

The frequency at 2159.20~$\mu Hz$ will be discussed in the section 4.4 alone.

  \subsection{Rotation splitting}

From the five multiplets of the l=1 modes, we derived an average rotation splitting of $1.07\pm0.05~\mu Hz$. Thus we estimate the rotation period of KUV~08368+4026 as $5.4\pm0.3$ days.

\begin{table}
\centering

\begin{tabular}{@{}lrrr}
\hline ID & $f$ & A & P \\
\hline \multicolumn{4}{c}{Independent signals}\\
\hline
 f1 & 1278.72 &  2.44 &  782.0 \\
 f2 & 1280.51 &  4.23 &  780.9 \\
 f3 & 1283.64 &  5.24 &  779.0 \\
 f4 & 1975.33 &  2.88 &  506.2 \\
 f5 & 1976.37 &  8.16 &  506.0 \\
 f6 & 1977.31 &  4.04 &  505.7 \\
 f7 & 2159.20 &  2.43 &  463.1 \\
 f8 & 2498.82 & 10.65 &  400.2 \\
 f9 & 2500.79 &  8.01 &  399.9 \\
f10 & 2758.35 &  3.30 &  362.5 \\
f11 & 2760.22 &  3.70 &  362.3 \\
f12 & 3245.61 &  4.47 &  308.1 \\
f13 & 3246.56 & 14.09 &  308.0 \\
f14 & 3248.17 &  7.17 &  307.9 \\
f15 & 3885.85 & 11.64 &  257.3 \\
f16 & 3886.90 & 10.39 &  257.3 \\
f17 & 3887.96 & 11.52 &  257.2 \\
\hline \multicolumn{4}{c}{Further signals}\\
\hline
f18 &  872.54 &  2.08 & 1146.1 \\
f19 &  875.50 &  1.42 & 1142.2 \\
f20 & 1115.80 &  1.55 &  896.2 \\
f21 & 1454.68 &  2.70 &  687.4 \\
f22 & 2191.49 &  1.80 &  456.3 \\
f23 & 2514.12 &  2.67 &  397.8 \\
f24 & 2606.27 &  1.77 &  383.7 \\
f25 & 3249.32 &  4.24 &  307.8 \\
f26 & 3387.63 &  3.18 &  295.2 \\
f27 & 3952.66 &  2.22 &  253.0 \\
f28 & 5234.48 &  2.21 &  191.0 \\
f29 & 5735.79 &  1.54 &  174.3 \\
f30 & 6492.81 &  1.41 &  154.0 \\

\hline
\end{tabular}
\caption{Signals identified. $f$=frequency in $\mu$Hz. $A$=amplitude in mma. P=Period in second.}
\end{table}

\begin{table}
\centering

\begin{tabular}{@{}cccccc}
\hline \multicolumn{6}{c}{Table a}\\
\hline $f$ & $\delta f$ & P & $\delta k$ & m & DP \\
\hline
1975.33 & &  506.2 & & -1 & \\
 & 1.04 & & & & \\
1976.37 & &  506.0 & + 5 & 0 &  1.38\\
 & 0.94 & & & & \\
1977.31 & &  505.7 & & +1 & \\
2159.20 & & 463.1 &  +4 & 0? &  7.77\\
2498.82 & &  400.2 & & -1 & \\
 & 0.99 & & & & \\
\textit{2499.81} & &  400.0 & + 3 & 0 & -6.10\\
 & 0.98 & & & & \\
2500.79 & &  399.9 & & +1 & \\
2758.35 & &  362.5 & & -1 & \\
 & 0.94 & & & & \\
\textit{2759.29} & &  362.4 & + 2 & 0 &  5.52\\
 & 0.93 & & & & \\
2760.22 & &  362.3 & & +1 & \\
3245.61 & &  308.1 & & -1 & \\
 & 0.95 & & & & \\
3246.56 & &  308.0 & + 1 & 0 &  0.36\\
 & 1.61 & & & & \\
3248.17 & &  307.9 & & +1 & \\
3885.85 & &  257.3 & & -1 & \\
 & 1.05 & & & & \\
3886.90 & &  257.3 & + 0 & 0 & -1.15\\
 & 1.06 & & & & \\
3887.96 & &  257.2 & & +1 & \\

\hline
\end{tabular}
\end{table}

\begin{table}
\centering
\begin{tabular}{@{}cccc}
\hline \multicolumn{4}{c}{Table b}\\
\hline $f$ & $\delta f$ & P & m  \\
\hline
1278.72 &  & 782.0 & -1 \\
 & 1.79 &  & \\
1280.51 &  &780.9 &  0 \\
 & 3.13 & & \\
1283.64 &  &779.0 & +2 \\

\hline
\end{tabular}
\caption{Modes identification for all l=1 modes(table a) and l=2 modes(table b). $f$ is the frequencies in $\mu Hz$, $\delta f$ is the frequency separation in $\mu Hz$. $P$ is the period in second and DP is the residual between observing period and the linear fit of period of the five l=1 modes. For the two doublets, the frequency values of the undetected m=0 mode are estimated at the center of the frequency values of the $m=\pm 1$ modes.}
\end{table}
\subsection{Period spacing}
In Table 3, there are some frequencies which belong to neither doublets nor triplets with amplitudes close to the detection limit and appearing only once among the five observation seasons. We identify these frequencies as further signals in Table 4, except the one at 2159.20~$\mu Hz$, which has been detected in four seasons hence listed as an independent signal.

Table 5 lists the identified l=1 frequencies and l=2 modes in Table a and b, respectively. With the three l=1 and m=0 modes in the triplets, We made a linear fit for the three modes which give an average spacing of 49.7s. Figure 3 shows the fitting. We also plot the missing m=0 modes with the frequency values of the center of the frequency values of the $m=\pm 1$ modes. The single mode in 2159.20$\mu Hz$ is plotted on the Figure as well.

\begin{figure}
\resizebox{\hsize}{!}{\includegraphics[angle=90]{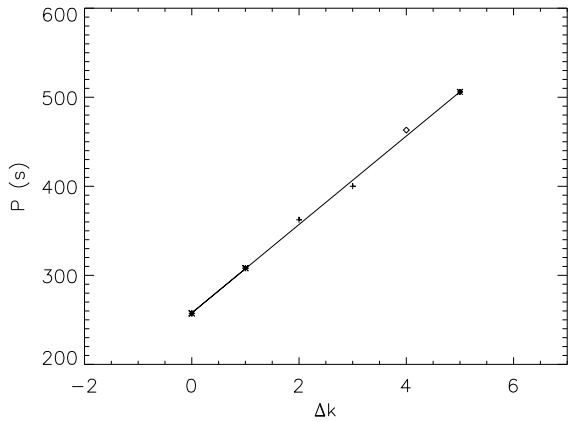}}
\caption{Linear fitting of three l=1, m=0 modes(star marks). The crosses show the two doublets, The circle shows the single mode of 2159.60~$\mu Hz$.}
\label{fig3}
\end{figure}

\subsection{Mode trapping}

\begin{figure}
\resizebox{\hsize}{!}{\includegraphics[angle=90]{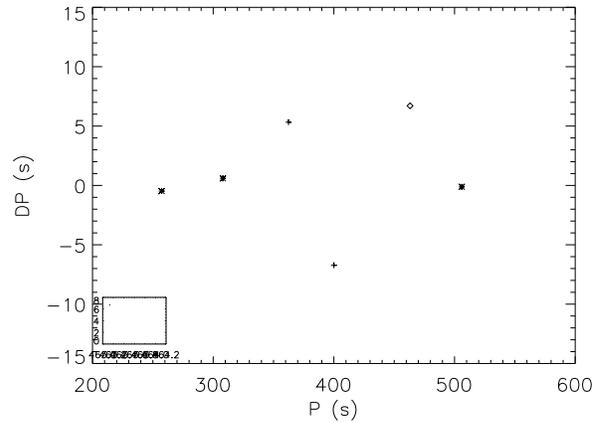}}
\caption{Residuals(DP) of the linear fit to the frequencies of 3 m=0 mode (star mark) versus the period. The residuals corresponding to the frequencies of the undetected m=0 modes in the two doublets are also plotted with crosses mark, the single mode point is shown with the open circle.}
\label{fig4}
\end{figure}
Figure 4 presents the residuals of linear fit for the frequencies of three m=0 modes, together with the residuals corresponding to the frequencies of the undetected m=0 modes in Table 5. The two doublets are also plotted, where the single mode point is shown with open circle. From Figure 4, a possible trapping mode could be visible for the period around 400s.

\section{Amplitude variations}
As mentioned before, KUV~08368+4026 shows a varying amplitude spectra in different observation seasons. The variations are in the scales of not only years but also weeks. Table 6 lists the amplitudes of the 25 frequencies resolved in dataset 3 in the three individual weeks from December 2009 to January 2010. Figure 5 shows the amplitude changes of each frequency in the three individual weeks. The amplitudes of these frequencies change a lot in a duration of around one month. We also calculate the total power for each week using the following equation:
\[
\rm{Total\, Power}=\sum_i A_i^2F_i
\]

Which are listed in the bottom of Table 6. As one may see, the total power of oscillation are changing in different weeks.

\begin{table}
\centering
\caption{The amplitudes list in the three weeks of Dataset 3. $f$ is the frequency in $\mu$Hz. $A$ is the amplitude in mma. Week1 is Dec.12-18, 2009, Week2 is Dec.26-31, 2009, Week3 is Jan.07-19, 2010. E is the total power of oscillation in $mma^2/s$}
\begin{tabular}{@{}lrrrr}
\hline  ID & $f$ &Week1&Week2&Week3\\
\cline{3-5}
  &  &  \multicolumn{3}{c}{A}\\
\hline
 c16 &   747.51 &  1.202 &  0.860 &  2.427 \\
 c22 &   875.50 &  1.686 &  0.343 &  1.846 \\
 c20 &  1127.55 &  0.821 &  1.465 &  1.730 \\
 c15 &  1270.98 &  3.770 &  1.804 &  1.731 \\
 c13 &  1387.15 &  2.352 &  2.346 &  2.304 \\
 c21 &  1975.43 &  3.123 &  4.394 &  2.929 \\
 c05 &  1976.44 &  8.510 &  5.013 &  7.637 \\
 c12 &  1977.42 &  3.442 &  1.616 &  3.216 \\
 c14 &  2159.25 &  1.594 &  1.242 &  2.882 \\
 c02 &  2498.91 & 12.749 & 17.280 & 11.541 \\
 c06 &  2500.62 &  4.455 &  2.898 &  6.990 \\
 c11 &  2758.44 &  2.838 &  4.389 &  2.222 \\
 c10 &  2760.33 &  1.427 &  2.022 &  3.892 \\
 c01 &  3246.39 & 11.886 & 18.218 & 16.682 \\
 c25 &  3247.29 &  7.266 &  6.284 &  2.792 \\
 c08 &  3248.56 &  6.243 &  6.943 &  4.699 \\
 c04 &  3886.01 & 11.633 &  9.866 & 11.345 \\
 c03 &  3887.13 &  8.581 & 15.438 &  9.120 \\
 c07 &  3888.02 &  9.229 & 13.032 & 11.119 \\
 c23 &  5747.48 &  1.218 &  0.812 &  1.858 \\
 c19 &  5863.48 &  2.051 &  1.013 &  1.529 \\
 c18 &  6385.99 &  0.964 &  1.827 &  1.877 \\
 c24 &  6492.81 &  0.491 &  0.891 &  1.930 \\
 c09 &  7134.27 &  3.904 &  3.391 &  3.222 \\
 c17 &  7774.06 &  1.822 &  1.150 &  1.985 \\
 & & & $E$ & \\
\cline{3-5}
 &  & 2.780 & 4.399 & 3.193 \\

\hline
\end{tabular}
\end{table}

\begin{figure}
\resizebox{\hsize}{!}{\includegraphics[angle=90]{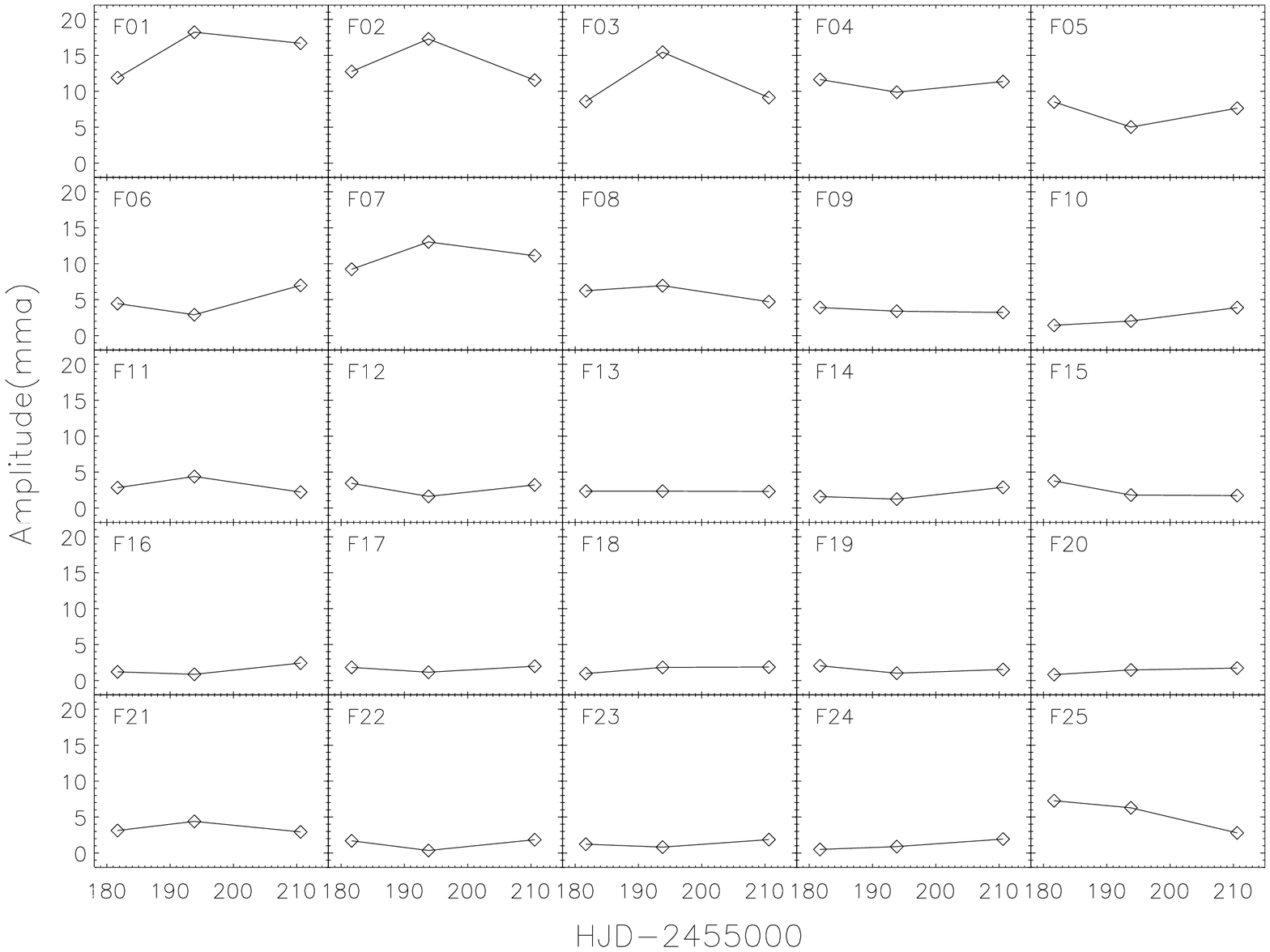}}
\caption{Amplitude variation of the 25 frequencies resolved in dataset 3 in different weeks.}
  \label{fig5}
\end{figure}

\section{Modeling exploration}
We try to use the theoretical static models calculated with the Toulouse white dwarf code (Pech, Vauclair \& Dolez 2006) to constrain the stellar parameters of KUV~08368+4026. The models have four input parameters, total mass, luminosity, hydrogen mass fraction and helium mass fraction. First we built a large grid in a large parameter range to select potential good-fit models. The ranges and steps of the grid are listed in Table 7. $\chi^2$ test estimates were used to find the best fit models with the five l=1 modes and the l=2 mode.

\[
\chi^2=\sum_n(P^{the}_n-P^{obs}_n)^2
\]
where $P^{the}$ are the periods of theoretical models and $P^{obs}$ are observing periods.

\begin{table}
\caption{Parameters of the large grid}
\begin{tabular}{@{}ccc}
\hline  & Range & Step \\
\hline Mass($M_{\sun}$) & $0.60\sim0.84$  & $0.01$\\
 Luminosity($L_{\sun}$) & $2.0\sim4.5\times 10^{-3}$ & $0.1\times 10^{-3}$\\
 $\log M_H/M_*$ &$-3.5\sim-10$&$0.5$\\
 $\log M_{He}/M_*$ &-2&\\
 \hline
\end{tabular}
\end{table}

More than 8000 models are calculated for the large grid. We constrain the parameters with the effective temperature and the surface gravity from Giannias catalog and Fontaine et al.(2003). We selected the models whose parameters locate inside the range of 3 times uncertainty of the parameters and one minimum of $\chi^2$ was found among the models. Around it we build a detailed grid to get more precise parameters. Table 8 lists the range of the detailed grid.

\begin{table}
\caption{Parameters of the detailed grid}
\begin{tabular}{@{}ccc}
\hline  & Ranges & Steps \\
\hline Mass($M_{\sun}$) & $0.67\sim0.71$  & $0.002$\\
 Luminosity($L_{\sun}$) & $2.7\sim3.1\times 10^{-3}$ & $0.02\times 10^{-3}$\\
 $\log M_H$ &$-4.0$&\\
 $\log M_{He}$ &-2&\\
 \hline
\end{tabular}
\end{table}

\begin{figure}
\centering
\resizebox{\hsize}{!}{\includegraphics{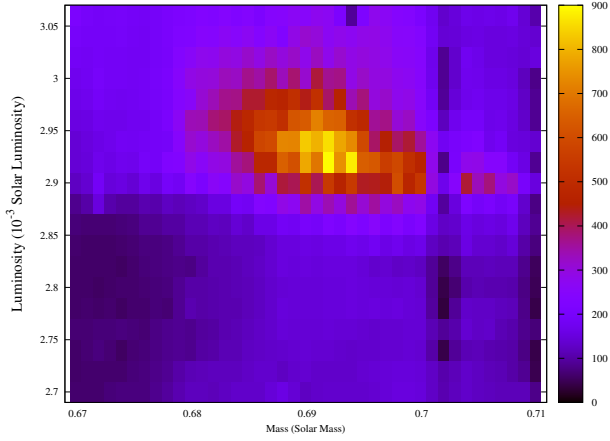}}
\caption{$\chi^2$ test of the detailed grids.}
    \label{fig7}
\end{figure}

Figure 6 displays the distribution of $\chi^2$ test of the eigen mode calculated from models and the observed mode. The $\chi^2$ values are represented by different gray scales. One minimum was found and the model was identified as the best fit model. Table 9 presents the parameters of the best fit model.

\begin{table}
\centering
\caption{Parameter of best fit model for KUV~08368+4026}
\begin{tabular}{@{}cc}
\hline $T_{eff}(K)$ &$11825.1$ \\
$\log g$ & $8.06$\\
$L(L_{\sun})$ & $(2.92\pm 0.02)\times10^{-3}$\\
$M_(M_{\sun})$ &$0.692\pm 0.002$ \\
$q_H$ &$10^{-4}$ \\
$q_{He}$&$10^{-2}$\\
\hline
\end{tabular}
\end{table}

We tried to do analysis for the trapping modes of the best fit model and the result is presented in Figure 7.

\begin{figure}
\centering
  \resizebox{\hsize}{!}{\includegraphics[angle=90]{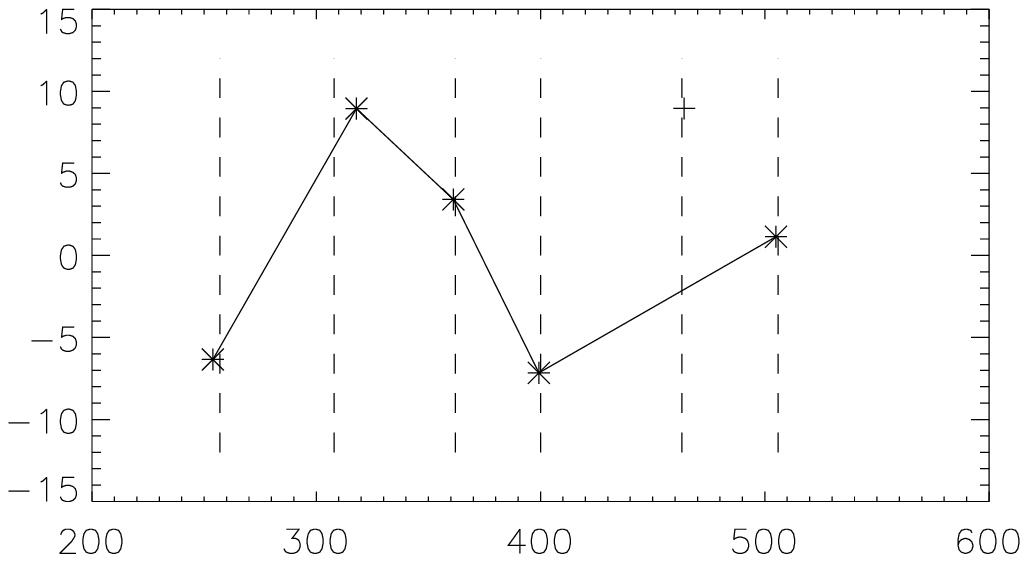}}
\caption{Mode trapping analysis of the seven best fit models. The symbols are the same as Fig.4.}
      \label{fig8}
\end{figure}

We suggest to take this model as the best fit model due to the following reasons: 1), it has the closest pulsation modes to the observation modes; 2), it shows a trapping mode around the 400s period, which agrees with the observation result well; 3), the effective temperature and surface gravity of this model locate between the values given by the two spectroscopic observation efforts. Therefore we take it as the best fit model under the current condition.

\begin{table}
\centering
\caption{Parameters of best fit model for KUV~08368+4026}
\begin{tabular}{@{}cc}
\hline $T_{eff}(K)$ &$11825.1$ \\
$\log g$ & $8.06$\\
$L(L_{\sun})$ & $(2.92\pm 0.02)\times10^{-3}$\\
$M(M_{\sun})$ &$0.692\pm 0.002$ \\
$q_H$ &$10^{-4}$ \\
$q_{He}$&$10^{-2}$\\
\hline
\end{tabular}
\end{table}

\section{Discussion and Conclusions}
A three-site observation campaign was carried out in 1998 (Dolez et al. 1999) when photelectric photometers were used. With the collected data, six independent modes were extracted while no multiplets detected. The Frequencies of the 6 modes agree with the central frequencies of the six multiplets detected in this work correspondingly.

We hence conclude our work as follows,

\begin{enumerate}
\item We obtained time-series photometric data for the ZZ Ceti star KUV~08368+4026 in 1999 and from 2009 to 2012. 17 independent modes were extracted, including six multiplets and one single mode together with 13 (f18-30) further signals. We identified the independent signals as either l=1 or l=2 modes with the rotation split. Also a number of linear combinations and low amplitude modes are resolved but we failed to identify them.\\
\item From the six multiplets, an average rotation split of $1.049\pm0.041\mu Hz$ was determined which derived the rotation period of $5.52\pm 0.22days$.\\
\item An average period spacing of 49.2s was obtained from the the l=1, m=0 modes.\\

\item All six multiplets were found in the 1998 campaign though the rotation splitting is not detected due to the observation conditions. The two periods of 619s and 494.5s found in the discovery observation were not detected in neither 1998 campaign nor the following observations.\\

\item We found the evidence of amplitude variations of KUV~08368+4026 in time scale of both years and weeks. The total pulsation power was changing as well in the 3 weeks from December 2009 to January 2010.\\

\item The theoretical modeling work suggests a thick hydrogen layer for KUV~08368+4026. We estimate a best fit model with the mass of $0.692\pm 0.002$ solar mass, luminosity of $(2.92\pm 0.02)\times 10^{-3}$ solar luminosity and hydrogen mass fraction of $10^{-4}$ stellar mass and helium mass fraction of $10^{-2}$ stellar mass\\
\item Romero et al. (2012) gave a group of stellar parameters from theoretical modeling work which suggested $log g$ of $8.02\pm0.03$, mass of $0.609\pm0.012$ solar mass, effective temperature of $11230\pm95$K, $M_H/M_*$ of $(1.42\pm0.52)\times10^{-5}$ and $M_{He}/M_*$ of $2.45\times 10^{-2}$ base on the two periods of the discovery data. Thus our constraints, which are base on more data of multisite campaigns should be considered to be more reliable. \\

\end{enumerate}

\section*{Acknowledgments}

CL and JNF acknowledge the support from the Joint Fund of Astronomy of National Natural Science Foundation of China (NSFC) and Chinese Academy of Sciences through the Grant U1231202, and the support from the National Basic Research Program of China (973 Program 2014CB845701 and 2013CB834904).

\label{lastpage}

\begin{thebibliography}{42}
\bibitem[\protect\citeauthoryear{Brickhill}{1975}]{1}Brickhill A. J., 1975, MNRAS, 170, 404
\bibitem[\protect\citeauthoryear{Dolez et al.}{1999}]{9}Dolez N., Vauclair G., Zhang X. B., Chevreton M., Handler G., 1999, ASPC, 169, 129
\bibitem[\protect\citeauthoryear{Fontaine et al.}{2003}]{13}Fontaine G., Bergeron P., Bill\`{e}res M., Charpinet S., 2003, ApJ, 591, 1184
\bibitem[\protect\citeauthoryear{Fu et al.}{2013}]{13}Fu J. -N. et al., 2013, MNRAS, 429, 1585*
\bibitem[\protect\citeauthoryear{Gianninas, Bergeron \& Ruiz}{2011}]{gcatalog}Gianninas A., Bergeron P., Ruiz M. T., 2011, ApJ, 743, 138
\bibitem[\protect\citeauthoryear{Landolt}{1968}]{41}Landolt A. U., 1968, ApJ, 153, 151
\bibitem[\protect\citeauthoryear{Lenz \& Breger}{2005}]{p04}Lenz P., Breger M., 2005, Comm. in Asteroseismology, 146, 53
\bibitem[\protect\citeauthoryear{Mukadam et al.}{2004}]{44}Mukadam A. S., Winget D. E., von Hippel T., Montgomery M. H., Kepler S. O., Costa A. F., 2004, ApJ, 612, 1052
\bibitem[\protect\citeauthoryear{Pech, Vauclair \& Dolez}{2006}]{42}Pech D., Vauclair G., Dolez N., 2006, A\&A, 446, 223
\bibitem[\protect\citeauthoryear{Romeo et al.}{2012}]{44}Romero A. D., G\'{o}rsico A. H., Althaus L. G., Kepler S. O., Castanheira B. G., Miller Bertolami M. M., 2012, MNRAS, 420, 1462*
\bibitem[\protect\citeauthoryear{Vauclair et al.}{1997}]{7}Vauclair G., Dolez N., Fu J. -N., Chevreton M., 1997, A\&A, 322, 155
\bibitem[\protect\citeauthoryear{Winget et al.}{1991}]{43}Winget D. E. et al., 1991, ApJ, 379, 326
\bibitem[\protect\citeauthoryear{Winget et al.}{1987}]{44}Winget D. E. et al., 1987, ApJ, 315, 77*
\end{thebibliography}
\end{document}